\documentclass[12pt,a4paper]{article}

\usepackage{jheppub}
\usepackage{comment}
\usepackage{graphicx}
\usepackage{amsmath}
\usepackage{dsfont}
\usepackage{slashed}
\usepackage{braket}
\usepackage{color}

\def\a{\alpha}
\def\ah{\hat\alpha}
\def\ta{\tilde{\alpha}}
 \def\p{\partial}
\def\ddp{\partial_+} \def\ddm{\partial_-}
\def\dda{\partial_a}
\def\ddau{\partial^a}
\def\L{\mathcal{L}}

\def\half{\frac{1}{2}}
      \def\ddp{\partial_+} \def\ddm{\partial_-}
    \def\dpY{\partial_{+}Y}  
     
\def\nn{\nonumber}

 \def\cd{\cdot}
\def\ep{e_+}
\def\emi{e_-}
\def\epm{e_+^{\mu}}
\def\emm{e_-^{\mu}}

\newcommand{\alpt}[1]{\tilde\alpha_{#1}}
\newcommand{\alp}[1]{\alpha_{#1}}
\newcommand{\alppl}[1]{\alpha_{#1,+}}
\newcommand{\alpmi}[1]{\alpha_{#1,-}}

\newcommand{\alptm}[1]{\tilde\alpha_{#1}^{\mu}}  \newcommand{\alptn}[1]{\tilde\alpha_{#1}^{\nu}}
\newcommand{\alpmh}[1]{\hat\alpha_{#1}^{\mu}}
\newcommand{\alpm}[1]{\alpha_{#1}^{\mu}}   \newcommand{\alpn}[1]{\alpha_{#1}^{\nu}}

\def\nn{\notag}    \def\ddo{\partial_1} \def\ddz{\partial_0}
\newcommand{\alignit}[1]{\begin{align}#1\end{align}}
\newcommand{\order}[1]{O(1/R^{#1})}
\def\ddz{\partial_0}
\def\ddo{\partial_1}
\def\Ym{Y^{\mu}}

\def\EOM{equations-of-motion}
\def\p{\partial}

\newcommand{\comments}[1]{}

\title{The effective string spectrum in the orthogonal gauge}

\author{Ofer Aharony, Matan Field and Nizan Klinghoffer\\ \it{Department of Particle Physics
and Astrophysics,\\Weizmann Institute of Science, Rehovot 76100, Israel} }
\emailAdd{Ofer.Aharony@weizmann.ac.il}
\emailAdd{Matan.Field@weizmann.ac.il} 
\emailAdd{Nizan.Klinghoffer@weizmann.ac.il}
\abstract{The low-energy effective action on long string-like objects in quantum field theory,
such as confining strings, includes the Nambu-Goto
action and then higher-derivative corrections. This action is diffeomorphism-invariant, and can be
analyzed in various gauges. Polchinski and Strominger suggested a specific way to analyze this
effective action in the orthogonal gauge, in which the induced metric on the worldsheet is conformally
equivalent to a flat metric. Their suggestion leads to a specific term at the next order beyond the Nambu-Goto action.
We compute the leading correction to the Nambu-Goto spectrum using the action that includes this term, and we show that it agrees with the leading correction previously computed in the static gauge. This gives a consistency check for the framework of Polchinski and Strominger, and helps to understand its relation to the theory in the static gauge.}

\begin{document}
\maketitle
 \bibliographystyle{JHEP}

\section{Introduction and summary of results}

Many field theories have stable extended string-like ($1+1$-dimensional) objects. Simple examples
include domain walls in $2+1$-dimensional field theories, ANO strings in the $3+1$-dimensional
Abelian Higgs model, and confining strings in non-Abelian gauge theories (the latter are not
stable in the presence of dynamical quarks, but are stable in pure Yang-Mills theory or in the
large $N$ limit). Consider the sector of the field theory which contains a single long string
(for instance, we can compactify the theory on a circle and wrap the string around the circle
to prevent it from shrinking). When the field theory has a mass gap, the low-energy effective
theory in such a sector contains only the massless excitations of the string. These always
include the transverse fluctuations of the string (which are classically Nambu-Goldstone bosons), and
generically these are the only massless fields on the string worldsheet. This low-energy effective
action is called the ``long string effective action.'' The effective theory has (at least) two relevant
scales, the tension $T$ of the string, and its length $R$ such that the characteristic excitation
energy of long strings is $1/R$. The effective action then has a derivative expansion in powers of
the dimensionless parameter $1 / T R^2$. As in many other contexts, this low-energy effective action is a
simple characteristic of the field theory which often captures interesting information about the
theory, so it is interesting to classify the possible terms that can appear in the effective
action, and to compute or measure them for various theories. The leading term in the low-energy
effective action is always the Nambu-Goto action (the area of the worldsheet multiplied by the
string tension), but generally this is corrected in various ways.

A priori there are no preferred coordinates on the long string worldsheet, so its action is
invariant under reparametrizations of these coordinates. There are two main gauge fixings that
have been used in the literature to study the effective action. One is the static gauge \cite{Luscher:1980fr,Luscher,LW},
where the worldsheet coordinates are chosen equal to two of the space-time coordinates. This
gauge is analogous to the unitarity gauge in spontaneously broken gauge theories; in it the
effective theory contains only the physical excitations of the string worldsheet.
The Lorentz symmetry of the underlying field theory is not manifest in this gauge, and can be
used to constrain the effective action \cite{Meyer,AK,AKS,AF}. The various allowed corrections to the Nambu-Goto
action in this gauge were analyzed in \cite{AK,AF,AD}, and the leading corrections to the Nambu-Goto values
for the long string energy levels were computed (and compared to lattice results for confining
strings) in \cite{ANK}.

A different gauge choice, the orthogonal gauge, in which the induced metric on the worldsheet is chosen to be conformally flat, was suggested twenty years ago by Polchinski and Strominger in
\cite{PS}, and analyzed further in \cite{Drum} (see also \cite{Natsuume:1992ky,HariDass:2006sd,Drummond:2006su,Dass:2006ud,HariDass:2007gn,DassMom}). In this gauge choice the symmetries are manifest, but
the action contains also ghosts and other non-physical degrees of freedom, as in Feynman gauge for
non-Abelian gauge theories. A rigorous analysis of the spectrum in this gauge seems complicated,
but Polchinski and Strominger suggested a specific way to quantize the effective string in this
gauge, which is analogous to the quantization of fundamental strings in the conformal gauge. Their
suggestion makes sense only when the leading correction to the Nambu-Goto action in the orthogonal
gauge has a specific coefficient.

Since there is no rigorous derivation of the suggestion of Polchinski and Strominger, it is not
clear if it is correct, and how it is related to the effective action in the static gauge. A naive
analysis indicates that if their suggestion is correct then the leading correction in the static
gauge (at least in $D > 3$ space-time dimensions) should be fixed to a specific value \cite{AKS}, but
it is not clear how to show this. In this paper we test the Polchinski-Strominger framework by
computing the leading corrections to the Nambu-Goto energy levels in this approach. The spectrum
at the leading orders in $1/TR^2$ was computed already in \cite{PS,Drum} and found to agree with the
Nambu-Goto spectrum, as found also in the static gauge approach in \cite{LW}. The leading deviation
from the Nambu-Goto energy levels in the static gauge approach occurs (for $D > 3$) at order
$1/R^5$, so in this paper we compute the energy levels in the Polchinski-Strominger approach
to this order. We find that (contrary to some claims in the literature \cite{HariDass:2009ub,HariDass:2009ud,Dass:2009xe}) there are
unique corrections to the Nambu-Goto spectrum at this order (for excited states in $D > 3$), and these corrections agree
with the known corrections in the static gauge approach, precisely when the coefficient of the
leading correction in that approach takes the value suggested in \cite{AKS}. This gives an important
consistency check for the framework of \cite{PS}, and also suggests that if this is the most
general possible framework, then the leading correction to the Nambu-Goto action (in any gauge)
is fixed, implying a sharp prediction (described in detail in \cite{ANK}) for the leading deviation of
the long string energy levels from their Nambu-Goto values. It would be very interesting to
compute directly from field theory (or from lattice simulations) the value of this leading
correction in some effective string theory, to see if it agrees with this prediction or not (this
will be discussed further in \cite{AKS}).

We begin in section \ref{sec:Formalism} by reviewing the framework of \cite{PS} and the resulting
computation of the spectrum of physical states at low orders in $1/R$. In section \ref{higher_orders} we extend this framework carefully to higher orders in $1/R$, and in section \ref{spectrum} we compute
the resulting spectrum, and compare it to the static gauge results.

\section{The framework and its leading order predictions}
\label{sec:Formalism}

In this section we review in detail the Polchinski-Strominger framework for analyzing the worldsheet theory of long strings, and the computation of its spectrum
at low orders in perturbation theory in the inverse length of the string. Most of these results have already appeared in \cite{PS,Drum}. As a preparation for the higher order computation we spend time on simplifying the formalism extensively. The next sections describe the generalization of these computations to higher orders.

\subsection{The classical theory}

The embedding of a string worldsheet into $D$-dimensional
flat space may be parameterized by $X^{\mu}(\sigma^a)$, ($\mu=0,1,...,D-1$, $a=0,1$), modulo reparametrizations (diffeomorphisms) of the worldsheet coordinates $\sigma^a$. We will assume that $X^{\mu}$ are the only massless fields on the string worldsheet. The reparametrization symmetry is gauged, so a gauge fixing needs to be chosen.
We will choose the \textit{orthogonal gauge}, in which the induced metric on the worldsheet is gauge fixed to be conformally flat (in Lorentzian signature\footnote{We work with space-time and worldsheet metric signatures $(-,+,+,...,+)$ and $(-,+)$, respectively.}),
\alignit{h_{ab}\equiv \partial_a X\cdot \partial_b X = e^{\phi}\eta_{ab}~,}
where $\cdot$ refers to a summation over space-time indices. In light-cone coordinates $\sigma^{\pm}=\sigma^0\pm\sigma^1$,
this gauge-fixing sets $h_{++}=h_{--}=0$, so the only non-vanishing component of the induced metric is
$Z\equiv h_{+-} = \partial_+ X \cdot \partial_- X$.

In this gauge the Nambu-Goto string action describing a string of tension $T$ is the free action\footnote{Note that the differential $d\sigma^+ d\sigma^-$ in our light-cone coordinates is twice $d^2\sigma$ in standard coordinates; this will contribute factors of $2$
in various places.}
\alignit{\label{eq:NG_action}
S_{NG}=T \int d^2\sigma \sqrt{-\det(h_{ab})} = T\int d\sigma^+d\sigma^- Z~.
}
\comments{
The quantization of this action gives the well-known spectrum of excitations of a long closed string of length $L$\cite{Arvis,...}
   \alignit{
  E_{N,\tilde N}= TL\sqrt{1+\frac{8\pi}{TL^2}\left(\frac{N+\tilde N}{2} -\frac{D-2}{24}\right)+\left(\frac{2\pi(N-\tilde N)}{TL^2}\right)^2}~,
 }
 where $N,\tilde N$ are the total excitation levels of the left and right movers, respectively. The spectrum is highly degenerate
since the energy depends only on the total excitation levels.
}
Polchinski and Strominger \cite{PS} argued that the leading correction to the Nambu-Goto action
in the orthogonal gauge, in
an expansion around a long string solution, is given by
\alignit{\label{eq:PS_action}
S_{PS}=\int d\sigma^+d\sigma^- \left[T Z+\frac{\beta}{4\pi}\frac{\ddp Z\ddm Z}{Z^2}\right]~,
}
and this will be our starting point.
The second term in \eqref{eq:PS_action} is singular for general configurations, but it is non-singular
in the derivative expansion around a long string solution that we will be working in.
Polchinski and Strominger noted \cite{PS} that this action can be heuristically 'derived' by taking Polyakov's effective action \cite{Pol} for the anomalous conformal factor in conformal gauge, $S_P=\frac{\beta}{4\pi}\int  d^2\sigma(\half\dda\phi\ddau\phi+\mu^2 e^{\phi})~,$ also known as the \textit{Liouville action}, and replacing the conformal factor of the intrinsic metric with the conformal factor of the induced metric, $e^\phi\rightarrow 2Z$, with $\mu^2\rightarrow4\pi T/\beta$.

The precise action appearing in \cite{PS} differs from \eqref{eq:PS_action} by terms proportional to the Nambu-Goto \EOM, so that their $X$'s are related to ours by a field redefinition,
but this difference will turn out to make no difference up to the order in $1/R$ that we work in.
Here we will work with \eqref{eq:PS_action}, in which the conformal transformation is the usual one. The action \eqref{eq:PS_action} is classically exactly conformal to all orders in our expansion; it is invariant under the transformations
 \alignit{\delta_{\pm}\sigma^{\pm}=\epsilon^{\pm}(\sigma^{\pm})~;~~\qquad\qquad ~~\delta_{\pm}X^{\mu}=\epsilon^{\pm}(\sigma^{\pm})\partial_{\pm}X^{\mu}~,}
 while the action in \cite{PS} is only conformal to leading order in $\beta$ (as also observed in \cite{HariDass:2007gn}).
By integration by parts, \eqref{eq:PS_action} can also be written as
\alignit{S_{PS}=\int d\sigma^+d\sigma^- \left[T Z+\frac{\beta}{4\pi}\frac{\ddp \ddm Z}{Z}\right]~.}

The variation of the action \eqref{eq:PS_action} is
\alignit{\delta S_{PS}=\int d^2\sigma \mathcal{J}\delta Z~,}
with
\alignit{\label{J}&\mathcal{J}\equiv T-\frac{\beta}{2\pi}\frac{\ddp\ddm \log(Z)}{Z}~.}
This gives the equations-of-motion
 \alignit{\label{eq:EOM}\ddm J_+^{\mu} + \ddp J_-^{\mu}=0~,}
which are simply the conservation equations for the space-time translation symmetry currents
\alignit{\label{Current}
 J_{\pm}^{\mu}&=\partial_{\pm}X^{\mu}\mathcal{J}~.}
The space-time momentum of the string is defined through this current\footnote{We consider
cylindrical string worldsheets, and choose the string spatial coordinate $\sigma^1$ to be
periodic with period $2\pi$.} :
\alignit{\label{STM}P^{\mu}=\int_0^{2\pi}d\sigma^1 J_0^{\mu}~.}

The energy-momentum tensor of the action \eqref{eq:PS_action} is
\alignit{\label{energy-momentum tensor}
T_{++}=2\pi h_{++} \mathcal{J} - \beta\mathcal{K_{++}}~,
}
with $\mathcal{J}$ defined in \eqref{J} and
\alignit{
&\mathcal{K_{++}}\equiv\partial_{+}^2 \log(Z) - \half (\partial_{+}\log(Z))^2~.}
A similar definition holds for the 'right-moving' part $T_{--}$, but since the whole analysis is completely identical we will focus all along on the 'left-moving' part.
From the equations-of-motion \eqref{eq:EOM} and the following identity
\alignit{\beta\ddm\mathcal{K}_{++}=-2\pi Z\ddp \mathcal{J}~,}
it follows that the energy-momentum tensor is conserved
\alignit{\ddm T_{++}=0~.}

 In order to have a situation with a stable long string, we consider a string wrapping a large compact dimension of radius $R$~,  $X^{\mu}(\sigma^0,\sigma^1+2\pi)\approx X^{\mu}(\sigma^0,\sigma^1)+2\pi R\delta_1^{\mu}$~.
 The simplest configuration
 \alignit{\label{eq:simple_solution} X^0=R\sigma^0~~,~~X^1=R\sigma^1~,}
 is a solution to the \EOM\ \eqref{eq:EOM}, and we will study the fluctuations around this solution.
 A more general solution to \eqref{eq:EOM} is
  \alignit{\label{aSolution}X^{\mu}_{cl}=R(\epm \sigma^++\emm\sigma^-)~,}
where in order to obey the gauge-fixing condition $h_{++}=h_{--}=0$, the two vectors should be null $\ep^2=\emi^2=0$,
 and for the periodicity they should also satisfy
 \alignit{\label{eq:constraint_on_e's}\epm-\emm=\delta_1^{\mu}~~\Rightarrow~~\ep\cd\emi=-\half~.}
Most of our computations will be done for general $e_{\pm}$, but we can always stick to \eqref{eq:simple_solution}, for which
 \alignit{\ep=\half(1,1,\vec 0)~~,~~\emi=\half(1,-1,\vec 0)~.}

 \subsubsection{Perturbation theory  in small fluctuations}

In the long string expansion, we expand
the fields (and then everything else) in powers of $1/R$ around the long string classical solution,
\alignit{\label{XtoY}
 X^{\mu}\equiv R(\epm\sigma^++\emm\sigma^-)+Y^{\mu}~,
 } where $Y^{\mu}$ are the fluctuations; the original symmetries will then be manifested non-linearly in the $Y$'s. In this expansion the equations-of-motion \eqref{eq:EOM} can be written as $\ddp\ddm\Ym=\order{3}~.$
Expanding the action with \eqref{XtoY} and defining
\alignit{W\equiv \ddp Y\cd\ddm Y~,~~M\equiv \ddp Y_-+\ddm Y_+~~~(Y_{\pm}\equiv Y\cd e_{\pm})~,}
we get (up to an additive constant)
\begin{equation}
\label{eq:LagLowerOrder}
 \L=TW-\frac{\beta}{\pi R^2}M\ddp\ddm M-\frac{2\beta}{\pi R^3}(W+M^2)\ddp\ddm M+\order{4}~.
\end{equation}

This action can be simplified significantly; when writing the general effective theory, terms which vanish when applying the free (zeroth order) equations-of-motion can be systematically ignored, since they can be canceled by field redefinitions, order by order. In our case we need to know the precise field redefinition that does the job; let us exemplify how this procedure works.
By integration by parts, every term that vanishes by the free equations-of-motion ($\ddp\ddm Y^{\mu}=0$) can be brought to the form
\alignit{\label{d+d-Y.F}\ddp\ddm Y\cd F[Y,\partial Y,...;e_{\pm}]~.}
Such a term is canceled when substituting the field redefinition $Y^{\mu}\rightarrow Y^{\mu}+\Delta Y^{\mu}$ into the free term $T\ddp Y\cd \ddm Y$ , if
\alignit{\Delta Y^{\mu}=\frac{F}{2T}~.}
Of course this field redefinition will lead to some other additional terms as well, but these would be at higher orders, and so this procedure can be maintained consistently order by order. Notice that the free equations-of-motion can be also written as $\ddp\ddm M=\order{3}~,$ and so the two higher-order terms in \eqref{eq:LagLowerOrder} can be eliminated in this way. For any $A[Y]$, integration by parts leads to
\alignit{\label{Ad+d-M->}A\ddp\ddm M\rightarrow -\ddp\ddm Y\cd \left[e_+\ddm A+c.c.\right]~,}
where 'c.c.' (``complex conjugate") means switching pluses and minuses everywhere.\comments{**in the terms sharing its parenthesis**.}
Then, when applying the field redefinition
 \alignit{\label{FRd3}\Delta Y^{\mu}=\frac{\beta}{2\pi TR^2}\left[e_+^{\mu}\ddm M+\frac{2}{R}e_+^{\mu}\ddm(W+M^2)+c.c.\right]~}
we end up with the free action \cite{Drum}
\alignit{\label{freeaction} \L=T\ddp Y\cd\ddm Y+\order{4}~,}
and free equations-of-motion
\alignit{\label{eq:EOM'}\ddp\ddm\Ym=\order{4}~,}
 at this order.

 The free action \eqref{freeaction} is the one we quantize. However, the energy-momentum tensor and space-time momentum need to be derived from the original action, and then to be transformed by the field redefinition \eqref{FRd3}.
Inserting the expansion \eqref{XtoY} into the space-time momentum \eqref{STM}, using the field redefinition \eqref{FRd3} and expanding in powers of $1/R$, we find
\alignit{\label{Momentum}
&P^{\mu}=\int_0^{2\pi}d\sigma^1\left\{TR\delta_0^{\mu}+T\partial_0 Y^{\mu}+\frac{\beta}{2\pi R^2}\left[\epm\ddm(\ddm+\ddz) M +c.c.\right]\right.\nn\\
&\left.\qquad\qquad\qquad+\frac{\beta}{\pi R^3}\left[\epm\ddm^2(W+M^2)+c.c.\right]
+\order{4} \right\}~.
 }
Noticing that
 \alignit{\partial_{\pm}^2=\pm \ddo\partial_{\pm}+\ddp\ddm~}
 and using the equations-of-motion \eqref{eq:EOM'} in $P^{\mu}$, we find the space-time momentum to be very simple \cite{DassMom} \footnote{The equations-of-motion can be used for our purpose in $P^{\mu}$, as well as in $T_{++}$, since we compute only their eigenvalues, and terms that are proportional to the equations-of-motion vanish between physical states.}
 \alignit{&P^{\mu}=2\pi TR\delta_0^{\mu}+T\int_0^{2\pi}d\sigma^1\ddz\Ym+\order{4}~.}\\
Similarly, putting \eqref{XtoY} into the energy-momentum tensor \eqref{energy-momentum tensor}, using the field redefinition \eqref{FRd3}, as well as the equations-of-motion \eqref{eq:EOM'}, the energy-momentum tensor is\footnote{This corrects a minus sign typo in \cite{Drum}.}
\begin{align}\label{EMT2}
T_{++}=&4\pi TR\ddp Y_+ + 2\pi T\dpY\cd\dpY+\frac{\beta}{R}\ddp^3 Y_- + \frac{\beta}{R^2}\ddp^2(\ddp Y_-)^2+\order{3}~.
\end{align}
The Virasoro generators are defined by
\alignit{\label{Virasoro_defined}
L_n=\frac{1}{2\pi}\int_0^{2\pi}d\sigma^1~T_{++}~ e^{+in\sigma^1}~.
}
Total $\sigma^1-$derivatives in $T_{++}$ do not contribute to $L_0$, which then simplifies to
\alignit{L_0=\int_0^{2\pi}d\sigma^1\left(2TR\ddp Y_+ + T\ddp Y\cd\ddp Y\right)+\order{3}~.}

\subsection{Quantization}

Upon quantizing the theory, the gauge-fixing procedure needs to be done carefully. The gauge-fixing
introduces a $(b,c)$ ghost system similar to the one which arises in the fundamental string in conformal gauge, and physical states may be identified with the cohomology of the BRST operator associated with
our gauge choice $h_{++}=h_{--}=0$. Unfortunately, this BRST operator is quite complicated, and it is
not clear how to work with it. Polchinski and Strominger suggested in \cite{PS} that the physical states
should be states obeying Virasoro constraints, of the same form as the ones arising in the conformal gauge
in ``old covariant quantization.'' This is the same as assuming that the BRST operator is equivalent
to the usual one for the fundamental string $Q_{PS} = \oint (c T_{++} + :b c \partial c: + {3\over 2} \partial^2 c)$. Such an assumption is consistent only if $Q_{PS}$ is nilpotent, which as usual requires
that $T_{++}$ \eqref{EMT2} satisfies a Virasoro algebra with central charge $c=26$, and Polchinski and Strominger showed that this is true if and only if $\beta = (26-D)/12$. In this paper we will follow
the assumptions of \cite{PS}, so we will take this value of $\beta$, and impose Virasoro constraints
$L_n \ket{\rm phys} = \delta_n^0 \ket{\rm phys}$ ($n \geq 0$) (and similarly for the right-movers) to define our physical states.

We quantize the theory using the canonical formalism in the Schr\"{o}dinger picture, where we expand the field operators and their conjugate momenta at a fixed time (say, $\sigma^0=0,~\sigma^1=\sigma$),
and impose the equal-time canonical commutation relations. This is different
from the previous approach of \cite{PS,Drum}, and it has the advantage of being more easily generalized to higher orders.
In particular, in this approach we will have fixed commutation relation for the modes, and higher order corrections will enter through the modification of the relation between the fields and their conjugate momenta.

\subsubsection{Mode expansions}

The fields $Y^{\mu}$ have conjugate momenta \alignit{\label{ConjugateMomentum}\Pi^{\mu}=\frac{\partial\L}{\partial\left(\ddz Y_{\mu}\right)}=T\ddz Y^{\mu} +\order{4}~.}
We expand the fields and their momenta at a fixed time\footnote{Everywhere an implicit summation is over all integers, and $\alpm{0}=\alptm{0}$.}
\alignit{\label{ModeExpansion}
Y^{\mu}(\sigma)&=\frac{i}{\sqrt{4\pi T}}\sum_{n\neq 0}\frac{1}{n}(\alpm{n}-\alptm{-n})e^{-in\sigma}+y^{\mu}~,\nn\\
\Pi^{\mu}(\sigma)&=\sqrt{\frac{T}{4\pi}}\sum_{n}(\alpm{n}+\alptm{-n})e^{-in\sigma}~.
}
Notice that these are two independent expansions.
 We demand canonical equal-time commutation relations
\alignit{[Y^{\mu}(\sigma),\Pi^{\nu}(\sigma')]&=i\eta^{\mu\nu}\delta(\sigma-\sigma')~,}
that lead to the usual Heisenberg algebra for the oscillators
\alignit{[\alpm{n},\alpn{m}]= [\alptm{n},\alptn{m}]=n\delta_{n+m}\eta^{\mu\nu}~;~~[\alpm{n},\alptn{m}]=0~.
}
From these mode expansions we compute all $\sigma$-derivatives of $Y$ and $\Pi$, and by using \eqref{ConjugateMomentum}
 and the equations-of-motion $\ddz^2 Y=\ddo^2 Y+\order{4}$ (which are used everywhere),
we can compute also $\tau$-derivatives and find
\alignit{\label{GeneralModeExpansions}
&\ddm^{k+1} Y^{\mu}=a\sum_{n} (-in)^{k} \alptm{n}e^{+in\sigma}+\order{4}~,\nn\\
&\ddp^{k+1} Y^{\mu}=a\sum_{n} (-in)^{k} \alpm{n}e^{-in\sigma}+\order{4}~,
} where we define $a\equiv 1/\sqrt{4\pi T}$ .

Putting the mode expansions \eqref{GeneralModeExpansions} into the energy-momentum tensor we find the Virasoro generators to be \cite{Drum}
\begin{align}\label{lns}
L_n=&\frac{R}{a}\alppl{n}+\half \sum_q:\alp{n-q}\cd\alp{q}:+\frac{\beta}{2}\delta_{n,0}\nn \\
&-\frac{\beta a n^2}{R}\alpmi{n}-\frac{\beta a^2 n^2}{R^2}\sum_q:\alpmi{n-q}\alpmi{q}:+\order{3}~,
\end{align}
where $\alp{n,\pm}\equiv \alp{n}\cd e_{\pm}$.
 Specifically,
\alignit{\label{lzero}
&L_0=\frac{R}{a}\alppl{0}+\half\alp{0}^2+\sum_{q>0}\alp{-q}\cd\alp{q}+\frac{\beta}{2}+\order{3}~.
}
Inserting the mode expansions \eqref{GeneralModeExpansions} into the space-time momentum \eqref{Momentum} we find
\alignit{\label{SimpleP}P^{\mu}=\frac{1}{a}\left(\frac{R}{2a}\delta_0^{\mu}+\alpmh{0}\right)+\order{4}~,}
where we distinguish the operator $\alpmh{0}$ from its eigenvalue $\alpm{0}$. When $\beta$ takes the value mentioned above,
\alignit{\label{forbeta} \beta=\frac{26-D}{12}~,}
one can show that the central charge in the Virasoro algebra of the $L_n$'s \eqref{lns} is equal to $c=26$ \cite{PS,Drum}.

The operators above are normal ordered, and the ordering constant ($a^m=\beta/2$ in \eqref{lzero}) can be obtained in several ways; additionally, the ordering constant
from the ghost system needs to be considered. It is important to emphasize that the ordering constant for each separate sector is not unique, and only the sum of ordering constants from the matter and ghost systems is well defined. Thus, it is important to extract the ordering constant by the same method for the matter and ghost systems. The commonly used physical state condition $L_n=\delta_{n,0}$ that we wrote above assumes the ghost ordering constant $a^g=-1$, and this value is obtained, for example, by the use of the Virasoro algebra, together with the mode algebra, or by considering the global Virasoro subalgebra between zero momentum ground-states. Different values ($a^m=-\frac{D}{24}$ and $a^g=\frac{1}{12}$, still leading to $a^m+a^g=\frac{2-D}{24}=\frac{\beta}{2}-1$) are obtained by taking a symmetric Weyl ordering and using zeta-function regularization for the infinite sums. In fact, in any method one is using at some point a regularization (and that is why only the final sum is meaningful).\\

\subsubsection{The spectrum}

The ground state of the long string (or the vacuum from the two-dimensional point of view) is the string state with no excitations and only space-time momentum
\alignit{\ket{0}\equiv\ket{0;k^{\mu}}~,} where $k^{\mu}$ is the eigenvalue of $P^{\mu}$.
  A general state then is obtained from the vacuum
 \alignit{\label{eq:all_states}
 (v_1\cd\alp{-n_1})\cdots (v_p\cd\alp{-n_p})(\tilde v_1\cd\alpt{-\tilde n_1})\cdots(\tilde v_q\cd\alpt{-\tilde n_q})\ket{0;k}~,}
 with total occupation numbers
 \alignit{N=\sum_{i=1}^p n_i~,~\tilde N=\sum_{j=1}^q \tilde n_j~.}
Translation invariance along the worldsheet relates $N-{\tilde N}$ to the longitudinal momentum $P^1$.

 We are looking for the string energy levels, and we focus on a string with vanishing transverse momentum; the complete spectrum is obtained from the static one by a boost. Thus we look for the eigenvalues of $P^0$, while demanding $P_{\perp}=0~.$ From \eqref{SimpleP}, fixing $P_{\perp}=0$ means $\alpha_{0,\perp}=0$, and we focus here on states with no longitudinal momentum ($P^1=0$),\footnote{These include the lowest states that get a contribution from the leading correction to Nambu-Goto in the static gauge \cite{ANK}.} which then satisfy the left-right level matching condition ($N=\tilde N$). For these states we can write
\alignit{&\alpmh{0}=\ah\delta_0^{\mu}~,}
and
\alignit{\label{SimplestP}P^0=\frac{1}{a}\left(\frac{R}{2a}+\ah\right)+\order{4}~.}
Demanding the zeroth constraints on the ground state
\alignit{\label{eq:zeroth_Virasoro_constraint}(L_0-1)\ket{0}=(\tilde L_0-1)\ket{0}=\order{3}~,}
we find an equation for $\alpha$
\alignit{\label{QuadraticEqForEpsilon}\half\alpha^2+\frac{R}{2a}\alpha-\frac{\beta-2}{2}+\order{3}=0~,}
which is solved by
\alignit{\alpha=-\frac{R}{2a}\pm\frac{R}{2a}\sqrt{1+\frac{8a^2}{R^2}\left(\frac{\beta-2}{2}\right)+\order{5}}~.}
We choose the positive physical frequencies,
and with \eqref{SimplestP} we finally get (using \eqref{forbeta})
 \alignit{E_0=\frac{R}{2a^2}\sqrt{1-\frac{8a^2}{R^2}\left(\frac{D-2}{24}\right)}+\order{4}~.}
  This is the familiar result for the Nambu-Goto ground-state energy, coming from a naive
  light-cone quantization of the Nambu-Goto action (as in \cite{Arvis}), usually written with the string tension $T=1/4\pi a^2$, and string length $L=2\pi R$ .


 Similarly, for a general state with no longitudinal momentum ($N=\tilde N$) the spectrum is easily shown to also agree with that \comments{Nambu-Goto spectrum (using \eqref{forbeta})}of the Nambu-Goto theory under naive light-cone quantization \cite{Arvis}:
   \alignit{
  E_N=\frac{R}{2a^2}\sqrt{1+\frac{8a^2}{R^2}\left(N-\frac{D-2}{24}\right)}+\order{4}~.
 }
 States with longitudinal momentum can be simply analyzed as well, but we will not do so here. It should be mentioned that although it is not manifest in our computation, the energy levels can only get contributions within an expansion in odd powers of $1/R$, as we indeed find. This is due to the symmetry of the action and its classical solution $R\rightarrow -R,~X^{\mu}\rightarrow -X^{\mu},~Y^{\mu}\rightarrow -Y^{\mu}$~, under which $E\rightarrow -E$.

\subsubsection{Physical Polarizations}

In general, for the states at level $N$, only the lowest $N$ Virasoro constraints are non-trivial. For the ground-state, which is
a scalar, there are no additional constraints; for the first level we need to further demand the first constraint to hold. Demanding
\alignit{L_1\ket{1_v;1_{\tilde v}}=0~,}
 where $\ket{1_v;1_{\tilde v}}\equiv (v\cd\alp{-1}) (\tilde v\cd\alpt{-1})\ket{0;k}$,
we find the physical state condition
\alignit{\label{eq:PhysicalityCondition}v_s\cd v=0~,}
with
\alignit{v_s^{\mu}= \left(\frac{R}{a}+\alpha\right)e_+^{\mu} + \left(\alpha-\frac{\beta a}{R}\left(1-\frac{a\alpha}{R}\right)\right)e_-^{\mu}+\order{3}~;}
this eliminates one degree of freedom in $v$. There is a similar condition on $\tilde v$, replacing pluses with minuses. In addition, there is a spurious state,
\alignit{\label{SpuriousState}L_{-1}\ket{0;1_{\tilde v}}=v_s\cd \alp{-1}\ket{0;1_{\tilde v}}=\ket{1_{v_s};1_{\tilde v}}~.}
This spurious state is physical if and only if
\alignit{\label{PhysicalSpurious}
v_s^2 &=-\left(\frac{R}{a}+\alpha\right)\left(\alpha\left(1+\frac{a^2\beta}{R^2}\right)-\frac{\beta a}{R}\right)=\order{2}~,\nn\\
\alpha &=-\frac{R}{2a}+\frac{R}{2a}\sqrt{1+\frac{4a^2\beta}{R^2}}+\order{4}~,}
 which means that the state has zero norm (as expected from a physical spurious state which is null).
 Indeed, the two equations \eqref{PhysicalSpurious} are consistent with each other,
up to the required order.
Every two states that differ by the null state \eqref{SpuriousState} are identified, and by that another degree of freedom in $v$ is eliminated; it is easily seen that the remaining $(D-2)$ physical polarizations are precisely the transverse ones.

This can in fact be generalized to all states in the theory, but in general
there are some corrections to the transverse states. Every physical state in the theory (using the
$L_n$'s up to $\order{2}$ as constructed above) can be written as a transverse state plus corrections
of order $1/R$ involving also longitudinal oscillators, and every transverse state can be corrected
in this way to give a physical state. The proof of this statement is a straightforward generalization
of the discussion in section 4.4 of Polchinski's book \cite{PolBook}; compared to the discussion there, we need to
exchange $+$'s with $-$'s, and the expansion of the BRST operator in eigenstates of $N^{lc}$ contains
also $Q_{-2}$, but precisely the same arguments show the one-to-one mapping between the transverse
states and the cohomology of $Q_{PS}$, and between the latter and the states of ``old covariant
quantization.'' Moreover, the generalization of equation (4.4.19) in \cite{PolBook} gives an
expansion whose leading term is the transverse state, and whose higher order terms are all
suppressed by at least one power of $1/R$.

To summarize, we have reproduced the fact \cite{Drum} that up to $\order{4}$ the Polchinski-Strominger string is iso-spectral to the Nambu-Goto string, and its states are in one-to-one correspondence with the excitations of the transverse polarizations to the world-sheet.

\section{Higher order corrections to the worldsheet charges}
\label{higher_orders}

In this section we extend the computation of $P^{\mu}$ and $T_{++}$ from section \ref{sec:Formalism} by two additional 
orders, that is up to $\order{5}$ for $P^{\mu}$ and up to $\order{4}$ for $T_{++}$. To do this we first extend the action, 
field redefinition and conjugate momentum, to $\order{5}$. At this order the corrections come solely from 
second term of \eqref{eq:PS_action}; all other possible terms in the effective action were shown to contribute only at 
higher orders in \cite{Drum}. 

\subsection{Action and field redefinition}

We start by considering the action up to $\order{5}$. 
The next two terms in the long-string expansion of the
action (\ref{eq:LagLowerOrder}) are :
\alignit{
&\L^{(4)}=-\frac{\beta}{\pi R^4}\left[\left(2M^2+W\right)\ddp\ddm W+4\left(M^3+MW\right)\ddp\ddm M\right]~,\nn\\
&\L^{(5)}=-\frac{2\beta}{\pi R^5}\left[2\left(M^3+MW\right)\ddp\ddm W+\left(W^2+6M^2W+4M^4\right)\ddp\ddm M\right]~.
}
There are additional contributions to this action which come from
the lower-order field redefinitions \eqref{FRd3}.
These new pieces will be proportional to
$\beta^2$,
\begin{align}
 \L_{\beta^2}^{(4)} & =  \frac{3 \beta^2}{4\pi^2 T R^4}  (\ddp \ddm M)^2~,\nn\\
\L_{\beta^2}^{(5)} & =  \frac{\beta^2}{\pi^2 T R^5}  \ddp \ddm M
\left[ \ddp \ddm \left( W+M^2\right) - \ddp^2 M \ddm Y_- +\half M\ddp\ddm M + c.c. \right].
\end{align}
Notice that by integration by parts we can take for any $A[Y]$
\alignit{\label{Ad+d-W->}A\ddp\ddm W\rightarrow A\ddp^2 Y\cd\ddm^2 Y-\ddp\ddm Y\cd \left[\ddm(\ddp Y A)-\half A\ddp\ddm Y+c.c.\right]~.}

The above action is quite complicated, but in a similar manner to the previous orders we can simplify it considerably. Using \eqref{Ad+d-M->} and \eqref{Ad+d-W->} we can bring many terms in the above action to the form \eqref{d+d-Y.F},
and performing the field redefinitions leaves us with the action
\begin{equation}
\label{eq:action}
 \L =  T W -\frac{\beta}{\pi R^4} \ddp^2 Y \cdot \ddm^2 Y \left(W + 2M^2 \right)  -
  \frac{4 \beta}{\pi R^5}\ddp^2 Y \cdot \ddm^2 Y \left(M^3 + M W \right)+\order{6}~.
\end{equation}
The field redefinitions at orders $\order{4,5}$ are
\begin{align}
\label{eq:frd}
\Delta Y^{\mu}_{(4)}  &=  \frac{\beta}{2\pi T R^{4}} \Big[ 4 e_{+}^{\mu} \p_{-}\left(M^{3}+MW\right)+ \p_{+}Y^{\mu}\p_{-}\left(W+2M^2\right) +c.c.\Big]~, \nonumber \\
\Delta Y^{\mu}_{(5)}  &=  \frac{\beta}{\pi T R^{5}}\Big[ e_{+}^{\mu}\p_{-} \left(W^{2}+6M^{2}W+4M^{4}\right)+ 2\p_{+}Y^{\mu}\p_{-}\left(M^{3}+MW\right) +c.c. \Big]\nn\\
&~-\frac{\beta^2}{2\pi^2 T^2R^5} \Big[e_+^{\mu} \p_-\Big(\p_{+}\p_{-}\left(W+M^{2}\right)-\p_{-}^{2}M\p_{+}Y_{+} +c.c \Big)+c.c. \Big]~.
\end{align}
Note that we have ignored terms that are proportional to the equations-of-motion inside the field redefinition, as these terms will not contribute to the energy levels.

\subsection{Conjugate momentum}

While at lower orders canonical quantization was straightforward, in our working order there are two complications.
The first is that the action \eqref{eq:action} contains second-order temporal derivatives of $Y$; the second is that
we will also have to take into account higher order corrections to the conjugate momentum.
To treat the temporal derivatives we apply another field redefinition, which will push
these terms beyond our working order.
Note that for any $A[Y]$,
\alignit{A\ddp^2 Y\cd\ddm^2 Y= -A\ddp\ddo Y\cd\ddm\ddo Y+A\ddp\ddm Y\cd\left(\ddo^2Y+\ddp\ddm Y\right),}
and on the right-hand side the higher time derivatives appear only in terms proportional to
the equations of motion.
Modifying our field redefinition accordingly leaves us with the final action
\begin{equation}
\label{eq:action2}
 \L =  T W +\frac{\beta}{\pi R^4}  \ddp\ddo Y\cd \ddm \ddo Y \left(W + 2M^2\right) +
  \frac{4 \beta}{\pi R^5}\ddp \p_1 Y \cdot \ddm \p_1  Y \left(M^3 + M W \right)+\order{6}~,
\end{equation}
which contains only first-order time derivatives. The revised field redefinitions are:
\begin{align}\label{eq:full_frd}
\Delta Y^{\mu}_{(4)} & =  \frac{\beta}{2\pi T R^{4}} \Big[ 4 e_{+}^{\mu}\p_-\left(M^3+MW\right)+\ddp Y^{\mu}\ddm\left(W+2M^2\right)-\half\ddo^2 Y^{\mu}\left(W+2M^2\right)+c.c.\Big]~, \nonumber \\
\Delta Y^{\mu}_{(5)} & = \frac{\beta}{\pi T R^{5}} \Big[ e_{+}^{\mu} \p_- \left(  W^{2}+6M^{2}W+4M^{4}\right)+ 2  \p_{+}Y^{\mu}\p_- \left(M^{3}+MW\right)-\ddo^2 Y^{\mu}\left(M^3+MW\right)  +c.c.\Big]\nn\\
&~-\frac{ \beta^2}{2 \pi^2 T^2R^5} \Big[e_+^{\mu}\p_-\Big(\p_{+}\p_{-}\left(W+M^{2}\right)-\p_{-}^{2}M\p_{+}Y_{+} +c.c. \Big) +c.c.\Big] ~.
\end{align}

The corrections to the conjugate momentum follow from its definition \eqref{ConjugateMomentum},
\begin{align}
\label{eq:momentum}
 \Pi^\mu &= T \p_0 Y^\mu+\Pi^\mu_{(4)}+\Pi^\mu_{(5)}+\order{6}~,\\
 \Pi^\mu_{(4)} & = \frac{\beta}{\pi R^4}\Big[\ddp\ddo Y\cd\ddm\ddo Y \Big(4M\delta_0^{\mu}+\ddz Y^{\mu}\Big)+\ddo\Big(\ddz\ddo Y^{\mu}\left(W+2M^2 \right)\Big)\Big] ~,\nn \\
\Pi^\mu_{(5)} & =  \frac{4\beta}{\pi R^5}\Big[\ddp\ddo Y\cd\ddm\ddo Y \Big( M \p_0 Y^\mu+\left( 3 M^2+W\right) \delta^{\mu}_0  \Big)+
 \p_1 \Big(\p_0 \p_1 Y^\mu \left( M^3 +M W\right) \Big)\Big]~.\nn
\end{align}

\subsection{Virasoro generators and space-time momentum}

To apply the constraints on the Hilbert space
we need to expand the original energy momentum tensor
to our working order, and then to apply to it the field redefinition. Applying the full field redefinition (\ref{eq:full_frd}) to $T_{++}$,
and then writing it in terms of canonical variables ($Y,\Pi$) with the help of (\ref{eq:momentum}), we find the corrections to \eqref{EMT2}:
\begin{align}\label{eq:full_emt}
 T^{(3)}_{++} & =  \frac{2 \beta}{ R^3} \p_1 \bigg[
-\frac{2}{3}\p_{1}M^{3}+W\p_{+}M-2M\p_{-}W+\p_{+}Y_-\p_{-}W\bigg] ~, \\
T^{(4)}_{++} & =  \frac{2 \beta}{ R^4} \bigg[  \p_{1}^{2}\Big[2WM^{2}-\frac{1}{2}W^{2}\Big]+\p_{1}\Big[4M\p_{-}^{2}Y_- \left(M^{2}+W\right)\Big]+\ddp\ddm W\left(W+2M^{2}\right)\bigg]~.\nn
\end{align}
Note that for convenience we write \eqref{eq:full_emt} in terms only of $Y$'s, even though it should be expressed in terms of $Y$'s and $\Pi$'s; that means that wherever there is a $\ddz Y^{\mu}$, it should be
replaced by a $\Pi^{\mu}/T$ (the corrections \eqref{eq:momentum} to $\Pi^{\mu} = T \ddz Y^{\mu}$ contribute to \eqref{eq:full_emt} at higher orders, which do not concern us). Notice also that any $\sigma^1$-derivatives will not contribute to $L_0$, and so will not affect the spectrum. We have verified the (on-shell) conservation of the energy-momentum tensor order-by-order. From \eqref{Virasoro_defined} and \eqref{eq:full_emt} we get: 
\begin{align}
\label{eq:Virasoro}
 L_0^{(3)}&= 0 ~,\\
L_0^{(4)}&=\frac{2\beta a^4 }{R^4}\sum_{p,q,k}pq\alp{p}\cd\alpt{q}\Big[\alp{k}\cd\alpt{p-q+k}+2\alp{k,-}\alp{-p+q-k,-}+
2\alpt{-k,+}\alpt{p-q+k,+}+4\alp{k,-}\alpt{p-q+k,+}\Big]\nn ~.
\end{align}

To quantize the generators at this order we choose the Weyl ordering scheme, as in \cite{ANK}. This means resolving the ambiguity in the transition from a classical function of modes to a quantum function of operators in a symmetric fashion between creation and annihilation operators. This scheme, followed by normal ordering the operators and $\zeta$-regularizing the difference, gives us no normal-ordering constant for $L_0$ at these orders.

To calculate the space-time spectrum to $\order{5}$ we need to also expand the original space-time momentum \eqref{STM} to this order:
\begin{eqnarray}
 P^{\mu(4) } & = & -\frac{2 \beta}{\pi R^4} \int d\sigma^1 \bigg[ 2\delta_0^{\mu}\left(M\ddp\ddm W+\ddp\ddm(M^3+MW)\right)+\ddz Y^{\mu}\ddp\ddm\left(W+2M^2\right) \bigg]~, \nn\\
P^{\mu(5)} & = & -\frac{4 \beta}{\pi R^5} \int d\sigma^1 \bigg[\delta_0^{\mu}\left((W+3M^2)\ddp\ddm W+\ddp\ddm(2M^4+3M^2W+\half W^2)\right)\nn\\
&&+~\ddz Y^{\mu}\left(M\ddp\ddm W+\ddp\ddm(M^3+MW)\right) \bigg]~,
\end{eqnarray}
where, as usual, we did not keep here terms that are proportional to the leading equations-of-motion. Applying to $P^{\mu}$ the complete field redefinition \eqref{eq:full_frd}, and then writing it in terms
of canonical variables ($Y,\Pi$), we get that these two contributions exactly vanish. So, in terms of modes, the space-time momentum is still the free one at this order,
\begin{eqnarray}
\label{eq:STM}
 P^\mu & = & \frac{1}{a} \left( \frac{R}{2a}\delta_0^{\mu}+\alpmh{0} \right) +\order{6}~.
\end{eqnarray}

\section{The spectrum and its comparison with the static gauge}
\label{spectrum}

We have seen that the Polchinski-Strominger string has the same energy spectrum as the Nambu-Goto string up to $\order{3}$. In this section we extend this result to order $\order{4}$, and find the first deviation from the Nambu-Goto spectrum at $\order{5}$. This deviation is found to match exactly with the first expected deviation from the Nambu-Goto spectrum in the static gauge, for a specific coefficient of the
leading deviation term $c_4$ there, thus supplying solid evidence for the equivalence of the two formalisms.

\subsection{The origin of corrections to the string spectrum}

The energy operator \eqref{eq:STM} is unchanged to $\order{5}$, so the energies are still
\begin{align} \label{energy}
\bra{n}P^0\ket{n}= \bra{n}\frac{R}{2a^2}+\frac{\ah}{a}\ket{n}=E_n~,
\end{align}
and the energy corrections come only from corrections to the Virasoro constraint. This corrects the value of $\alpha$, which is determined (as a function of $R$) by
 \begin{align}\label{eq:highest_Virasoro}
\bra{n}L_0^{[0]}+L_0^{(4)}-1\ket{n}=\order{5}~,
\end{align}
with $L_0^{[0]}$ the Virasoro operator computed in section \ref{sec:Formalism},
 \alignit{
 L_0^{[0]}&\equiv \frac{R}{a}\alppl{0}+\half\alp{0}^2+\sum_{q>0}\alp{-q}\cd\alp{q}+\frac{\beta}{2}~,
 }
 and $L_0^{(4)}$ given by \eqref{eq:Virasoro} (recall $L_0^{(3)}=0$).
  Note that the physical states $\ket{n}$ are defined to be eigenstates both of the Virasoro operator $L_0$ (with eigenvalue one) and of the space-time momentum $P^{\mu}$ (where for our states only $P^0$ is non-zero).

  We have two kinds of corrections in the Virasoro equation \eqref{eq:highest_Virasoro}, relative to \eqref{eq:zeroth_Virasoro_constraint}; there is the correction to the Virasoro operator at order $\order{4}$, $L_0^{(4)}$, and there is also the correction to the string states
\begin{align}
\ket{n}=\ket{n^{(0)}_{\alpha}}+\ket{\Delta n}~,
\end{align}
where $\ket{n^{(0)}_{\alpha}}$ are the free transverse states \eqref{eq:all_states}
with a momentum related to $\alpha$ by \eqref{eq:STM}
(these are eigenstates of $L_0^{[0]}$ for any $\alpha$),
and $\ket{\Delta n}$ are the corrections. Since the correction to the Virasoro generators $L_n$ is $\order{3}$, we
do not have to turn on $\ket{\Delta n}$ until this order.
Expanding \eqref{eq:highest_Virasoro},
\begin{align}
\bra{n^{(0)}_{\alpha}+\Delta n}L_0^{[0]}+L_0^{(4)}-1\ket{n^{(0)}_{\alpha}+\Delta n}=\order{5}~,
\end{align}
we see that $\ket{n^{(0)}_{\alpha}}$ should satisfy $(L_0^{[0]}-1)\ket{n^{(0)}_{\alpha}}=\order{4}~,$ and then
the corrections to the string states $\ket{\Delta n}$ do not contribute to this equation to the
order that we are working in. Thus, to determine $\alpha$ to our working order, we can evaluate
the Virasoro constraint between free eigenstates
\begin{align}\label{eq:between_free_eigenstates}
\bra{n^{(0)}_{\alpha}}L_0^{[0]}+L_0^{(4)}-1\ket{n^{(0)}_{\alpha}}=\order{5}~.
\end{align}
Even better, in \eqref{eq:between_free_eigenstates} it is enough to consider only the terms $L_{0,diag}^{(4)}$ in $L_0^{(4)}$ that preserve the excitation levels of the states, separately for the left-movers and right-movers.
 The other terms, while modifying the physical states, do not correct the eigenvalues (and thus the energy spectrum) at leading order in the perturbing operator; this is a standard feature of first order in perturbation theory. So finally, to obtain the energy levels to $\order{5}$, we need to find $\alpha$
 such that
\begin{align} \label{final_eq}
\bra{n^{(0)}_{\alpha}}L_0^{[0]}+L_{0,diag}^{(4)}-1\ket{n^{(0)}_{\alpha}}=\order{5}~.
\end{align}

\subsection{The correction to the spectrum}

Looking at the  correction to the Virasoro operator \eqref{eq:Virasoro}, we consider (as explained above) only its level-preserving components
\begin{align}
\label{eq:L4}
L_{0,diag}^{(4)}= \frac{2\beta a^4}{R^4}\sum_{n,m}nm~ \alp{n}\cd\alpt{m}\Big[\alp{-n}\cd\alpt{-m}+4\alp{-n,-}\alpt{-m,+}\Big]~.
\end{align}
Working with the Weyl scheme, normal ordering the operators and regularizing the infinite sums, we can write
\begin{align}
\label{eq:L4b}
L_{0,diag}^{(4)}= \frac{2\beta a^4}{R^4}\left(\Sigma^{\mu\nu}\tilde \Sigma_{\mu\nu}+4\Sigma^{\mu}_{~-}\tilde \Sigma_{\mu+}\right)~,
\end{align}
with the left-movers part $\Sigma^{\mu \nu}$ defined as
\begin{align}\label{eq:Sigma}
\Sigma^{\mu\nu}&\equiv \sum_{n=1}^{\infty}n\left(\alpm{-n}\alpn{n}-\alpn{-n}\alpm{n}\right)~,
\end{align}
and a similar right-movers part, with $\a$'s replaced by $\tilde\a$'s. Noting that in our conventions the $\pm$ indices are lowered with $-1/2\big(\begin{smallmatrix}0&1\\1&0\end{smallmatrix}\big)$, \eqref{eq:L4b} can be rewritten as
 \begin{align}\label{eq:L4c}
L_{0,diag}^{(4)}=\frac{2\beta a^4}{R^4}\left(\Sigma^{ij}\tilde \Sigma_{ij}+\Sigma^{\mu-}\tilde \Sigma_{\mu-}-\Sigma^{\mu+}\tilde \Sigma_{\mu+}\right)~,
\end{align}
where $i,j$ are summed over the $(D-2)$ transverse dimensions.
From the discussion above it is clear that only states with \comments{equal left and right excitation levels ($N=\tilde N$)}both left and right excitations get contributions from this correction; in particular \comments{and that}the ground state (annihilated by $\Sigma^{\mu \nu}$) has no correction to its energy at this order, similarly to the situation in the static gauge \cite{ANK}. The first corrected states then have $\ket{N, \tilde N} =\ket{1,\tilde 1}$.
We showed above that at the leading order the physical $\ket{1,\tilde 1}$ states are the same as
the transverse states, and we
can decompose the transverse $\ket{1,\tilde 1}$ states into $SO(D-2)$ representations:
\begin{align}
& \ket{1,\tilde 1;\mathds{1}} \equiv \vec\a_{-1}\cd\vec\ta_{-1}\ket{0}~,\nn\\
 &\ket{1,\tilde 1; (i,j)} \equiv \left( \a_{-1}^{(i} \ta_{-1}^{ j)}
-\frac{2}{D-2}\delta^{ij} \vec\a_{-1}\cd\vec\ta_{-1}\right) \ket{0}~,\nn\\
& \ket{1,\tilde 1; [i,j]}  \equiv \a_{-1}^{[i}\ta_{-1}^{j]} \ket{0}~,
\end{align}
where we use $\vec\alpha\cd\vec{\tilde\alpha}\equiv\sum_i\alpha^i\cd\tilde\alpha^i$ for a summation only over transverse indices. Considering first \comments{at}the transverse part of \eqref{eq:L4c} acting on a general $\ket{1,\tilde 1}$ state, we find
\begin{align}
\Sigma^{ij}\tilde\Sigma_{ij}\a_{-1}^k\tilde\a_{-1}^l\ket{0}=\Big[-2\a_{-1}^l\tilde\a_{-1}^k+
2\delta^{kl}\vec\a_{-1}\cd\vec\ta_{-1}\Big]\ket{0}~.
\end{align}
This leads to:
\begin{align}
&\Sigma^{ij}\tilde\Sigma_{ij}\ket{1,\tilde 1;\mathds{1}}=2(D-3)\ket{1,\tilde 1;\mathds{1}}~,\nn\\
&\Sigma^{ij}\tilde\Sigma_{ij}\ket{1,\tilde 1;(ij)}=-2\ket{1,\tilde 1;(ij)}~,\nn\\
&\Sigma^{ij}\tilde\Sigma_{ij}\ket{1,\tilde 1;[ij]}=2\ket{1,\tilde 1;[ij]}~.
\end{align}
The non-transverse part of \eqref{eq:L4c} vanishes between any transverse states,
\begin{align}
\bra{n_{trans}}\Sigma^{\mu\pm}\ket{n_{trans}}=0~,
\end{align}
 as can be seen by acting on a single mode,
 \begin{align}
 \Sigma^{\mu\pm}\a_{-1}^k\ket{0}=\left(\delta^{k\pm}\a_{-1}^{\mu}-\delta^{\mu k}\a_{-1}^-\right)\ket{0}~,
 \end{align}
 and closing from the left with another transverse state.
We then get the eigenvalues of $L_{0,diag}^{(4)}$ when acting on $\ket{1,\tilde 1}$ states
\begin{align}
L_{0,diag}^{(4)}\ket{1,\tilde 1;i}=\lambda_i\frac{4\beta a^4}{R^4} \ket{1,\tilde 1}~,
\end{align}
with $\lambda_i=(D-3)~,-1~,+1$ for the scalar, symmetric and anti-symmetric states, respectively ($i=1,2,3$).
The energy levels depend on $\a$
\begin{align}
E=\frac{1}{a}\left(\frac{R}{2a}+\a\right)+\order{6}~,
\end{align}
 which is fixed, accordingly to \eqref{eq:between_free_eigenstates}, by the corrected equation
 \begin{align}
 \half\a^2+\frac{R}{2a}\a-\frac{\beta}{2}+\lambda_i\frac{4\beta a^4}{R^4}=\order{5}~.
 \end{align}
The energy levels of the various $\ket{1,\tilde 1}$ states are then
\begin{align}
E_{1,1;i}=\frac{R}{2a^2}\sqrt{1+\frac{4a^2\beta}{R^2}-\lambda_i \frac{32\beta a^6}{R^6}}+\order{6}~,
\end{align}
from which the leading correction to the Nambu-Goto spectrum is found for the various $SO(D-2)$ representations of $\ket{1,\tilde 1}$ states:
\begin{align}\label{eq:DeltaE_OG}
\Delta E_{1,1}=-\frac{8\beta a^4}{R^5}
\left\{\begin{tabular}{cc}
D-3&\mbox{scalar}\\
-1&\mbox{symmetric}\\
+1&\mbox{anti-symmetric}\\
\end{tabular}\right\}~,
\end{align}
with $\beta=(26-D)/12$ .

From the discussion above it is clear that the only contribution to \eqref{final_eq} when
the states are transverse comes from the $\Sigma^{ij} {\tilde \Sigma}_{ij}$ term in $L_0^{(4)}$;
but since physical states differ from transverse states only at $O(1/R)$, this implies that
the leading ($\order{5}$) correction to the energy of any physical
state is proportional to the eigenvalue of $\Sigma^{ij} {\tilde \Sigma}_{ij}$ on the transverse state
associated with that physical state.
In particular, since for
$D=3$ the transverse $\Sigma^{ij}$ vanish, there are no corrections to energy levels at $\order{5}$
for $D=3$ .

\subsection{Comparison to static gauge}

The general effective string action in static gauge in a derivative expansion, up to six-derivative order, was analyzed in \cite{AK}, and was found to coincide with the Nambu-Goto action, apart from a single allowed deviation,
\begin{align}
-c_4 \int d^2\sigma \left(\partial_a\partial_b \vec X\cd\partial^a\partial^b \vec X\right)\left(\partial_c \vec X\cd\partial^c \vec X\right)~,
\end{align}
where the implicit summation here is over the $(D-2)$ transverse space-time coordinates (which are the
only fields in the static gauge).
The leading correction to the Nambu-Goto energy levels, coming from this term in the action, was computed for the states with $N=\tilde{N}=1$ in \cite{ANK}, and was found to be (in the case of the closed string)
\begin{align}\label{eq:DeltaE_SG}
\Delta E_{1,1}=\frac{256\pi^4 c_4}{T^2L^5}
\left\{\begin{tabular}{cc}
D-3&\mbox{scalar}\\
-1&\mbox{symmetric}\\
+1&\mbox{anti-symmetric}\\
\end{tabular}\right\}~.
\end{align}
Comparing to \eqref{eq:DeltaE_OG}, using $T=1/4\pi a^2,L=2\pi R$ and the critical value of $\beta$,
we see that the corrections to the
energy levels that we found in the Polchinski-Strominger formalism are the same to $\order{5}$ as
those in the static gauge, when $c_4$ has the specific value
\begin{align} \label{forcfour}
c_4 = \frac{D-26}{192\pi}=-\frac{\beta}{16\pi}~.
\end{align}
It is easy to see that this agreement in fact extends to all states (to $\order{5}$), since the
static gauge correction was found in \cite{ANK} to be proportional to the same operator $\Sigma^{ij}
{\tilde \Sigma}_{ij}$ that we found above in the orthogonal gauge. The value \eqref{forcfour} was
previously conjectured to be related to the Polchinski-Strominger formalism in \cite{AKS}, and here
we showed that this conjecture is consistent with the spectrum to $\order{5}$ .

To summarize, we have computed the leading correction -- at $\order{5}$ -- to the Nambu-Goto energy levels
in the orthogonal gauge approach of Polchinski and Strominger, and found that it precisely agrees with
the correction previously computed in the static gauge approach, for the specific value \eqref{forcfour}
of the coefficient of the leading correction to the Nambu-Goto action in that approach. It would be
interesting to extend this computation to higher orders in $1/R$, taking into account that at
higher orders quantum corrections to the Polchinski-Strominger action \eqref{eq:PS_action} may be required to
preserve conformal invariance.

\section*{Acknowledgements}

 We would like to thank
 Z. Komargodski and A. Schwimmer for many useful discussions and for comments on a draft of this paper. OA would like to thank J. Drummond and J. Polchinski for useful discussions. This work was supported in part by the Israel--U.S.~Binational Science Foundation, by a research center supported by the Israel Science Foundation (grant number 1468/06), by the German-Israeli Foundation (GIF) for Scientific Research and Development, and by the Minerva foundation with funding from the Federal German Ministry for Education and Research.

%

\bibliography{PS}

\end{document}